\documentclass[aps,pre,twocolumn,superscriptaddress,amsmath,amssymb,showpacs,10pt]{revtex4-1}

\usepackage{graphicx}
\usepackage{epstopdf}
\usepackage{tabularx}
\usepackage{subfigure}
\usepackage{amsmath}
\usepackage{amsmath}
\usepackage{amstext}
\usepackage{amssymb}
\usepackage{color}

\begin{document}

%
\title{Granular chains with soft boundaries: slowing the transition to quasi-equilibrium.}

%
\author{Michelle Przedborski}
\email{mp06lj@brocku.ca}
\affiliation{Department of Physics, Brock University, St. Catharines, Ontario L2S 3A1, Canada}

\author{Thad A. Harroun}
\email{thad.harroun@brocku.ca}
\affiliation{Department of Physics, Brock University, St. Catharines, Ontario L2S 3A1, Canada}

\author{Surajit Sen}
\email{sen@buffalo.edu}
\affiliation{Department of Physics, State University of New York, Buffalo, New York 14260-1500, USA and Department of Physics, Brock University, St. Catharines, Ontario L2S 3A1, Canada}

\date{\today}

%
%
\begin{abstract}
We present here a detailed numerical study of the dynamical behaviour of `soft' uncompressed grains in a granular chain where the grains interact via the intrinsically nonlinear Hertz force. It is well known that such a chain supports the formation of solitary waves (SW). Here, however, the system response to the material properties of the grains and boundaries is further explored. In particular, we examine the details of the transition of the system from a SW phase to an equilibrium-like (or quasi-equilibrium) phase and for this reason we ignore the effects of dissipation in this study. We find that the soft walls slow the reflection of SWs at the boundaries of the system, which in turn slows the journey to quasi-equilibrium. Moreover, the increased grain-wall compression as the boundaries are softened results in fewer average grain-grain contacts at any given time in the quasi-equilibrium phase. These effects lead to increased kinetic energy fluctuations in the short term in softer systems. We conclude with a toy model which exploits the results of soft-wall systems. This toy model supports the formation of breather-like entities and may therefore be useful for localizing energy in desired places in the granular chain.  
\end{abstract}


\maketitle
%
%
%
\section{\label{sec:intro} Introduction}
Following the pioneering works of Nesterenko~\cite{Nesterenko1983,Nesterenko1985,Nesterenko1995,Sinkovits1995,Sen1996,Coste1997,Sen1998,Chatterjee1999,Hong1999,Hinch1999,Hascoeet2000,Sen2001,Sen2001b}, a tremendous amount of experimental, analytical, and numerical studies into the dynamics of granular chains have emerged in recent years (see e.g. Refs.~\onlinecite{Ji1999,Nesterenko2001,Rosas2003,Rosas2004,Nesterenko2005,Job2005,Sen2008,Santibanez2011}). It is well established that, in general, any velocity perturbation to an end grain in a 1D granular chain without any initial pre-compression, i.e. an unloaded chain, will propagate through the system as a nondispersive bundle of energy, or solitary wave (SW)~\cite{Sen2008}. This SW develops within approximately ten grain diameters~\cite{Sokolow2007}, and it travels through the chain with a constant speed which scales with SW amplitude, as is characteristic for nonlinear waves~\cite{Sen2008}. This SW also has a well-defined and fixed width that is sensitive to the grain shape~\cite{Sun2013}, and equal to approximately five grain diameters for grains with ellipsoidal contact geometries such as spheres~\cite{Nesterenko1983,Sen2001}.  


The SW arises from the contact potential between grains described by the Hertz law~\cite{Hertz1882}, $V \sim \delta^n$, where $\delta >0$ is the grain-grain overlap and $n=5/2$ for spherical grains. The intrinsically nonlinear Hertz potential possesses no quadratic term, and thus one cannot have harmonic motion in the unloaded granular chain. This manifests itself in the absence of acoustic propagation in the system~\cite{Vitelli2012}, a fact referred to by Nesterenko and others as the ``sonic vacuum''~\cite{Nesterenko1983}. Adding compression into the chain effectively introduces a harmonic term to the potential, and the effects of this term on the existence and stability of SWs has been extensively studied elsewhere, see e.g. Refs.~\onlinecite{Nesterenko2001} and \onlinecite{Takato2012}.

The SW supported by the granular chain is unlike solitons found in other (i.e. continuum) physical systems: it is a unique SW specific to discrete lattices with purely nonlinear interactions. The unloaded chain's discrete nature has the possibility of broken grain-grain contacts, and this crucial feature means that, in general, SWs are not perfectly preserved. Collisions of SWs with other SWs has been extensively probed in the literature, see e.g. Refs.~\onlinecite{Manciu1999},~\onlinecite{Avalos2007} and~\onlinecite{Avalos2011}. SWs in the granular chain are also not preserved in collisions with boundaries~\cite{Sen2004,Sen2005,Mohan2005,Job2005}. Rather, the SW breaks up and reforms in the collision process~\cite{Avalos2007}, resulting in the partial destruction of the original SW and the birth of secondary solitary waves (SSWs) which are much smaller in magnitude, but possess the same spatial extent as the parent SW~\cite{Manciu2002}. 

SSWs were first reported in numerical studies in Refs.~\onlinecite{Manciu2002} and~\onlinecite{Manciu2000} and later confirmed experimentally by Job et al.~\cite{Job2005}. It is now known that the presence of boundaries in these systems generally leads to the formation of  SSWs with various energies which end up moving at a distribution of velocities related to the energy carried in the SSW. In the absence of dissipation and at sufficiently long time after an initial energy perturbation to the system, the rates of breakdown and creation processes of SSWs balance and the chain reaches a steady state which has been referred to in the literature as the \textit{quasi-equilibrium} (QEQ) phase~\cite{Avalos2011,Sen2004,Sen2005,Mohan2005}. Here we investigate how the system's journey to QEQ can be tuned by varying the material parameters of the granular chain system, and the effects of introducing an inertial mismatch at the boundary on the onset and rate of relaxation to the QEQ phase. We will ignore the effects of dissipation that are invariably present in a system made up of macroscopic elastic objects. A dissipation free system should be regarded as a first and necessary step in reaching driven-dissipative systems with QEQ-like behavior. 

Here we extend the work of Ref.~\onlinecite{Avalos2014} by considering the effects of either softening the boundary walls or the whole system on the dynamics of the unloaded granular chain held between fixed symmetric boundaries. We are particularly concerned with the effects of soft boundaries on the rate of relaxation as the system slips into QEQ. As we show below, the time delay introduced by the soft walls in the rebounding of SWs at the boundaries increases with the wall softness. This time delay has two main effects on the system dynamics: in the short term, soft-wall systems exhibit larger kinetic energy fluctuations than corresponding systems with homogeneous or hard walls; in the long term, the transition to the QEQ phase is slowed down by incorporating soft boundaries into the system. We suggest that, in soft-wall systems, the journey from the initial SW state to the QEQ phase can be understood in terms of a transition from a SW state to an ergodic phase where the ensemble average of grain energy is equal to the time average. 

One key interest in the QEQ phase is with possible energy harvesting and localization applications. Recent interest in these systems has probed metastable special regions such as so-called discrete breathers~\cite{Sievers1988,James2013,Boechler2010} and hot and cold spots that are sustained for extended times~\cite{Avalos2011}, where the kinetic energy can be localized along well-defined locations.  Here we use insights from results of soft-wall systems to discuss a toy model which may be useful for locally trapping energy.

\section{ \label{sec:methods} Methods}
We consider $N$ monodisperse spherical grains in 1D, held between two fixed symmetric walls, under zero pre-compression. Initially the grains are barely touching, but as any two grains in the chain make contact they interact via a contact potential given by the Hertz law~\cite{Hertz1882}, $V(\delta_{i,i+1}) = a_{i,i+1} \delta_{i,i+1}^{5/2}$, where $\delta_{i,i+1} = (R_{i+1}-R_i) - (x_{i+1} - x_i)$ is the grain-grain overlap with $x_i$ denoting the absolute position of grain $i$ with respect to the origin, and $R_i$ its radius. The proportionality constant is given by
\begin{equation}
a_{i,i+1} = \frac{2}{5 D_{i,i+1}}\sqrt{\frac{R_i R_{i+1}}{R_i + R_{i+1}}}, 
\label{eq:Fa}
\end{equation}
where the prefactor $D_{i,i+1}$ is related to the Young's modulus $Y$ and Poisson's ratio $\sigma$ of the grains by
\begin{equation} 
 D_{i,i+1} = \frac{3}{4}\left[ \frac{1-\sigma_i^2}{Y_i} + \frac{1-\sigma_{i+1}^2}{Y_{i+1}} \right ].
\label{eq:D}
\end{equation}
We consider here chains in which all the grains are comprised of the same material, and thus  $\sigma_i = \sigma_{i+1}$ and $Y_i = Y_{i+1}~\forall~i$. It is then possible to express the equations of motion for any grain (except the end grains) in the chain as follows:
\begin{equation}
m\ddot{x}_i = \frac{5}{2}  \left [ a_{i-1,i} \delta_{i-1,i}^{3/2}-a_{i,i+1} \delta_{i,i+1}^{3/2} \right ],
\label{eq:motion}
\end{equation}
where $m$ is the grain mass, $\ddot{x}_i$ denotes the second derivative of the absolute position with respect to time, and $i=2\dots N-1$. To implement the fixed walls, one takes $R_j \to \infty$ as the radius of the wall, ensuring that the boundary is much larger than any grain in the chain and will therefore not move during the simulation, while simultaneously relaxing the condition that the wall must be flat. In this way, the potential that develops at the interface between either the first or the last grain in the chain and the wall is given by: 
$V = (2/5D_{i,j})\sqrt{R}\delta_{i,j}^{5/2}$, where $R$ is the grain radius for the end grains.

In this work we refer to a chain in which the grains and walls are comprised of the same material as a \textit{homogeneous} chain. One requires only a single force prefactor to fully characterize the system in this case, since $D_{i,i+1} = D_{i,j} \equiv D$. The potential that develops between an end grain and wall is then a factor of $\sqrt{2}$ larger than the grain-grain potential in the homogeneous systems. The chains in which the grain material differs from the wall material are referred to in this work as \textit{inhomogeneous} systems. In this case, one requires two force prefactors to characterize the system: one for the grain-grain interactions, $D_{i,i+1} \equiv D_g$, and one for the grain-wall interactions $D_{i,j} \equiv D_\textrm{w}$. Clearly, $D_g = D_\textrm{w}$ in homogeneous granular systems.

We use the velocity-verlet algorithm~\cite{Allen1987} with a time step of $dt = 10^{-5}~\mu\mathrm{s}$ to integrate the equations of motion. Initially all of the grains are at rest, and the left end grain is given a velocity perturbation directed into the chain at $t = 0~\mu\mathrm{s}$ of magnitude $v_0 = 9.899\times10^{-5}~\mathrm{mm}/\mu\mathrm{s}$ in all cases. The system is then left to evolve in time according to the equations of motion. Typically, chains were comprised of $N=38$ spherical grains, and all grains were taken to be $R=6~\mathrm{mm}$ in radius. The choice of system size was linked to probing chains which may be experimentally accessible, and we show in Sec.~\ref{sec:transition} that this is a suitable choice. We considered a variety of materials for the homogeneous systems whose material parameters can be found in the appendix, sorted according to the product of the mass density $\rho$ and prefactor $D$ (vide infra). These materials range from as hard as  diamond ($\rho D=0.00415\,(\mathrm{mm}/\mu\mathrm{s})^{-2}$) to as soft as silicone rubber ($\rho D=52.4\:(\mathrm{mm}/\mu\mathrm{s})^{-2}$). 

For inhomogeneous systems, all chains were taken to be comprised of grains with $Y = 193~\mathrm{GPa}$, $\sigma = 0.305$, and $\rho = 7.82~\mathrm{mg/mm^3}$, consistent with material properties of stainless steel. This corresponds to a grain-grain force prefactor of $D_g = 0.0071$~$(\mathrm{mm}\,\mu\mathrm{s}^{2}/\mathrm{mg})$. Keeping the grain material fixed, we then symmetrically vary the wall material parameters to elucidate the effects of softening the boundaries on the system dynamics.

For the inhomogeneous systems, the $D_\textrm{w}$ values were chosen to vary over a suitable range, thus some are contrived and do not correspond to any particular material. We considered $D_\textrm{w} = 0.0041$ (diamond walls), 0.0071 (stainless steel walls), 0.05, 0.5, 5, 10, 19 ($\approx$ rubber walls), 30, 50, and 75~$(\mathrm{mm}\,\mu\mathrm{s}^{2}/\mathrm{mg})$.  A peculiar feature of these systems is that the notion of ``hard'' and ``soft'' walls is not absolute, but is relative to the material comprising the grains in the chain, and the walls can be considered soft only if $D_{\textrm{w}} > D_g$. For example, the stainless steel systems considered here begin to exhibit soft-wall behaviour only once $D_{\textrm{w}}> 0.0071$. 

\section{ \label{sec:results} Numerical results}
\begin{figure*}[htp]
\centering
\includegraphics[scale=0.8]{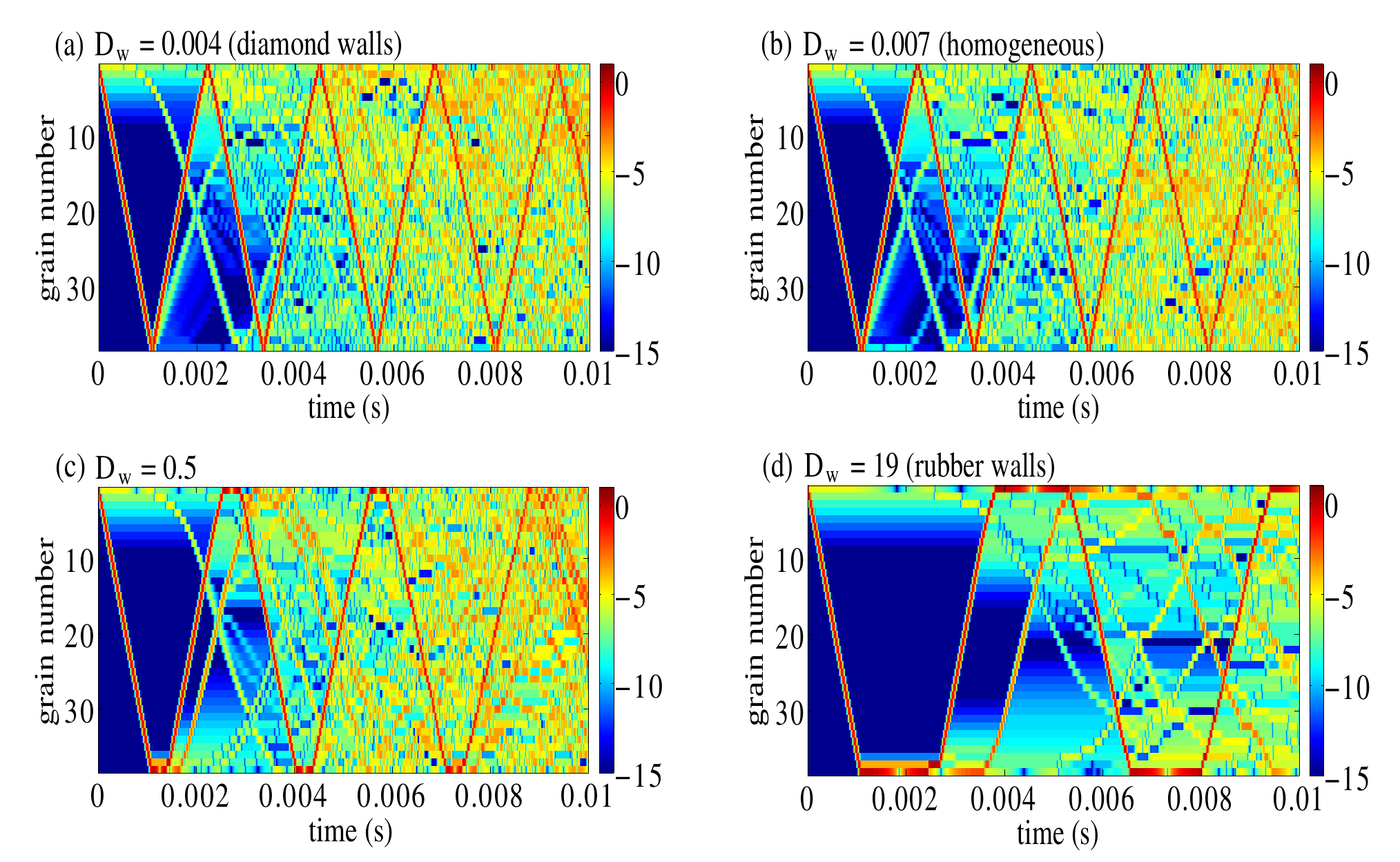} 
\caption{(Color online) Kinetic energy density plots for the $N=38$ \textit{inhomogeneous} stainless steel granular chains with varying wall softness $D_\textrm{w}$. For each panel (a) - (d), $D_g = 0.0071$. The kinetic energy values are normalized by the total system energy and presented using a logarithmic scale.  The SW speed is set by the grain material, see Fig.~\ref{fig:fig2}, and is the same for all systems presented here. Notice the progressive delay that develops in the rebounding of SWs at the boundaries as they are made increasingly softer.}
\label{fig:fig1}
\end{figure*}

\begin{figure*}[htp]
\centering
\includegraphics[scale=0.8]{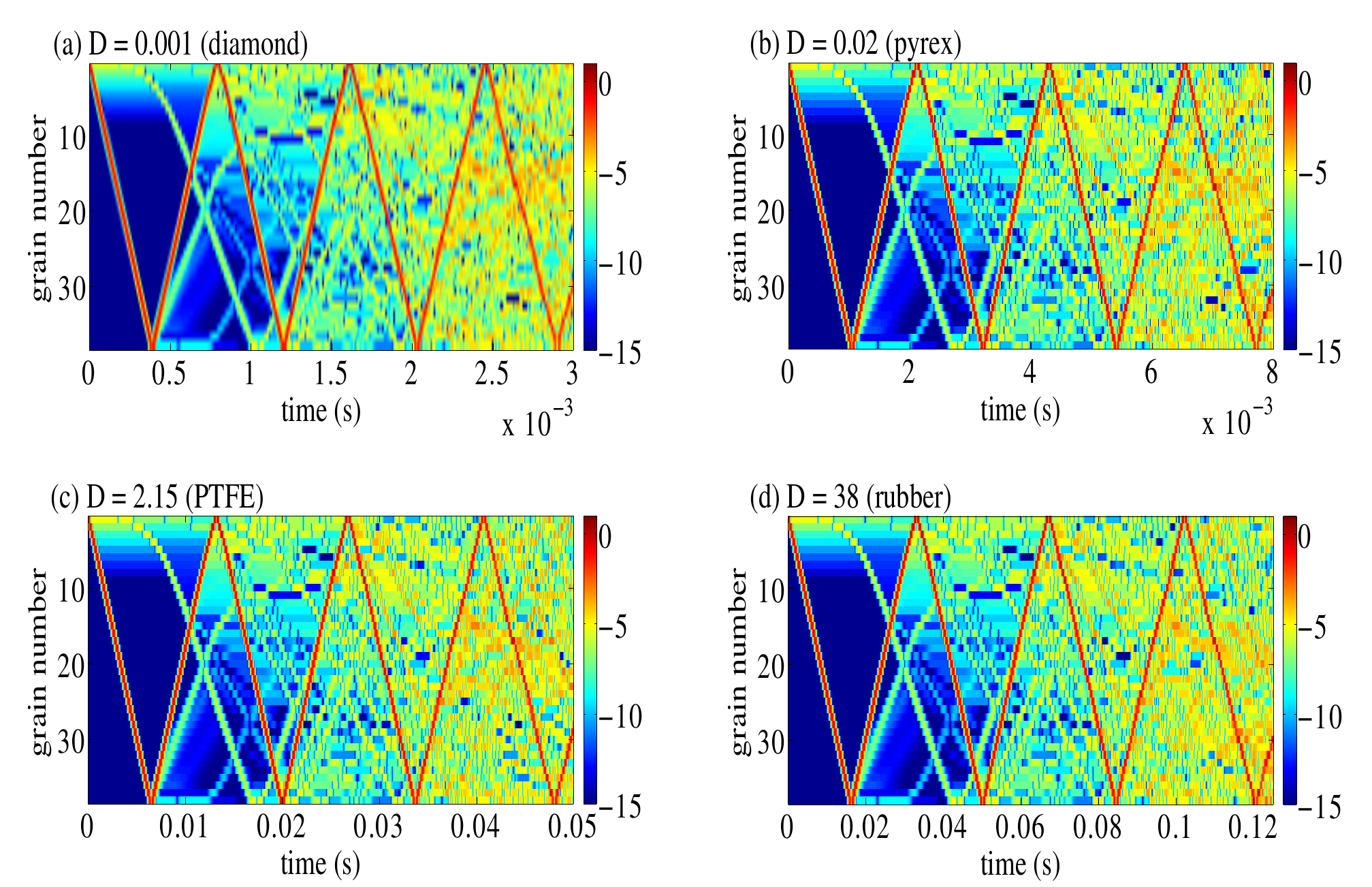} 
\caption{(Color online) Kinetic energy density plots for the \textit{homogeneous} $N=38$ granular chains comprised of varying material parameters. Kinetic energy values were normalized by the total system energy and presented on a logarithmic scale. The time scale has been adjusted to capture the first four rebounds of the ISW at the right wall. Recall for homogeneous systems $D_\textrm{w} = D_g = D$. The SW possesses a speed which is set by the softness of the grains, and progressively softer systems ($\rho D \gg 1$) correspond to smaller wave speeds.}
\label{fig:fig2}
\end{figure*}

After initiating a delta-function velocity perturbation to the left end grain at $t = 0~\mu\mathrm{s}$, it takes several grains for the SW to develop in the system, as illustrated by kinetic energy density plots in Figs.~\ref{fig:fig1} and \ref{fig:fig2}. A small amount of kinetic energy remains at the left wall due to the left end grain rebounding off its neighbour to transfer the kinetic energy of the initial solitary wave (ISW), down the chain. Likewise, after the ISW reaches and emerges from the right wall, secondary solitary waves (SSWs) are formed at the right wall from a similar rebounding of the right end grains. These energy processes occur in both the inhomogeneous and homogeneous systems and are visible for a variety of cases in the early time dynamics of Figs.~\ref{fig:fig1} and \ref{fig:fig2}.

The most striking feature of Fig.~\ref{fig:fig1} is the emergence of a time delay in the ISW rebounding at the boundary in the inhomogeneous systems as one progressively softens the walls. This appears as a horizontal line segment near grains 1 and 38, and is absent in the hard-wall and homogeneous systems. In contrast to homogeneous systems, soft wall systems take a notable amount of time ($\approx 1600~\mu\mathrm{s}$ for $D_\textrm{w}=19$) to rebound from the compression of the end grain as the ISW collides with the far wall. This slow time scale associated with interactions at the walls  and a much faster time scale associated with SW interactions within the chain (i.e. away from the wall) leads to an inertial mismatch between the grains and the walls, or the appearance of ``sticky'' walls in the inhomogeneous systems. One finds that the softer the wall is made to be, the longer the time it takes to re-emit the SW from the boundary. 

According to the virial theorem, a system with potential energy $V \sim \delta^{n}$ will partition the total system energy as $n/(n+2)$ kinetic energy and $2/(n+2)$ potential energy. For $n=5/2$, this leads to $5/9$ of the total energy as kinetic energy, and the remaining $4/9$ as potential energy. Most of the system's energy is held as potential energy while the end grains are compressed against the soft walls for extended periods of time. Thus the soft-wall systems make larger SSWs to compensate, in an attempt to partition the system's total energy as appropriate. This process is particularly visible by comparing kinetic energy density plots in Fig.~\ref{fig:fig1} for $D_\textrm{w} = 0.007$ (homogeneous) and $D_\textrm{w} = 19$: the SSW emitted at the right wall near $0.0025~\mathrm{s}$ in the inhomogeneous soft-wall system has a much larger energy content, thus appearing ``warmer'' than that emitted in the homogeneous case. 

Another interesting feature of Fig.~\ref{fig:fig1} are the ``bending waves'' which appear as curved line segments (e.g. beginning at about $0.004~\mathrm{s}$ for $D_\textrm{w} = 19$) and occur more prominently in the soft-wall systems than homogeneous systems. Any time grains break contact, a gap is formed in the system causing the SW to lose its coherence and partially break. Thus some fraction of the SW kinetic energy located in broken-contact grains becomes decoupled from the main part of the wave, forming new small SSWs. Propagation of a SSW is slowed by gaps between grains; until the grain with peak kinetic energy catches up with the rest of the chain, the progress of the SSW cannot continue. As inter-grain gaps are closed, the SSW picks up speed until reaching its natural velocity, which is determined by the Hertz potential exponent, wave amplitude, and grain material properties~\cite{Sen2008}. Thus many SSWs appear to `curve' in the kinetic energy density plot. We speculate that soft-wall systems facilitate the opening of larger gaps in the system, leading to more prominent bending of the SSWs.

An increased tendency to open larger and more frequent gaps in the system intuitively results from the soft walls being more easily compressed by the harder grains. In fact, the first and last grains always remain in contact with the soft walls during the rebounding of SWs and the emission of SSWs at the boundaries. Once the last grain initially makes contact with the right wall in a soft-wall system, it never breaks contact for the duration of the simulation. This is also true for the left end grain, after rebounding off the second grain. The sustained grain-wall compression facilitates the opening of gaps near the ends of the chain, which are evident from an analysis of individual grain kinetic energies over time. For the steel grain/rubber wall ($D_\textrm{w} = 19$) case for example, it takes a significant amount of time for grain 38 to pass its kinetic energy to grain 37, and also for 37 to pass to 36, etc. as the ISW is rebounding from the right wall. This is because the end grains have become disconnected from each other due to the opening of a noticeably large gap at the right end of the system as grain 38 remains compressed into the right wall. Consequently, end grains travel with a constant velocity before striking their neighbours: grain 38 travels $\approx 4 ~\mu\mathrm{m}$ before hitting grain 37, and grain 37 travels $\approx 10$ times this distance before colliding with grain 36 in this system.

\subsection{\label{sec:transition} Transition to quasi-equilibrium}
The build up of SSWs from inter-grain gaps and extended wall-grain contacts eventually destroys the phase dominated by the ISW, and the system transitions into a new phase with many small SSWs breaking and reforming very quickly. This can be seen in the emergence of noisy energy fluctuations in Fig.~\ref{fig:fig1} at longer times. 

This \textit{quasi}-equilibrium (QEQ) phase for the initial conditions used here is marked by a Maxwell-Boltzmann distribution of particle velocities, but where the equipartitioning of energy among all grains is absent. Since the ISW is partially destroyed in collisions with system boundaries- which, as a byproduct slowly produce SSWs in the vicinity of the collision region- the energy content of the ISW decreases over time, and energy is partially distributed among other grains in these unbiased systems. The motion of individual grains in one-dimensional systems is highly correlated, hence groups of particles often oscillate together. The fastest grain in the chain at any instant will either be part of the ISW, in which case it will have a large share of the system's kinetic energy, or it will have only a small share of the kinetic energy of several SSWs in the QEQ phase. Therefore to visualize the transition from the unstable phase made of various SWs and SW-like structures to the QEQ phase, one can consider simply the speed of the fastest grain in the chain. 

\begin{figure*}[htp]
\centering
\includegraphics[width=0.8\textwidth]{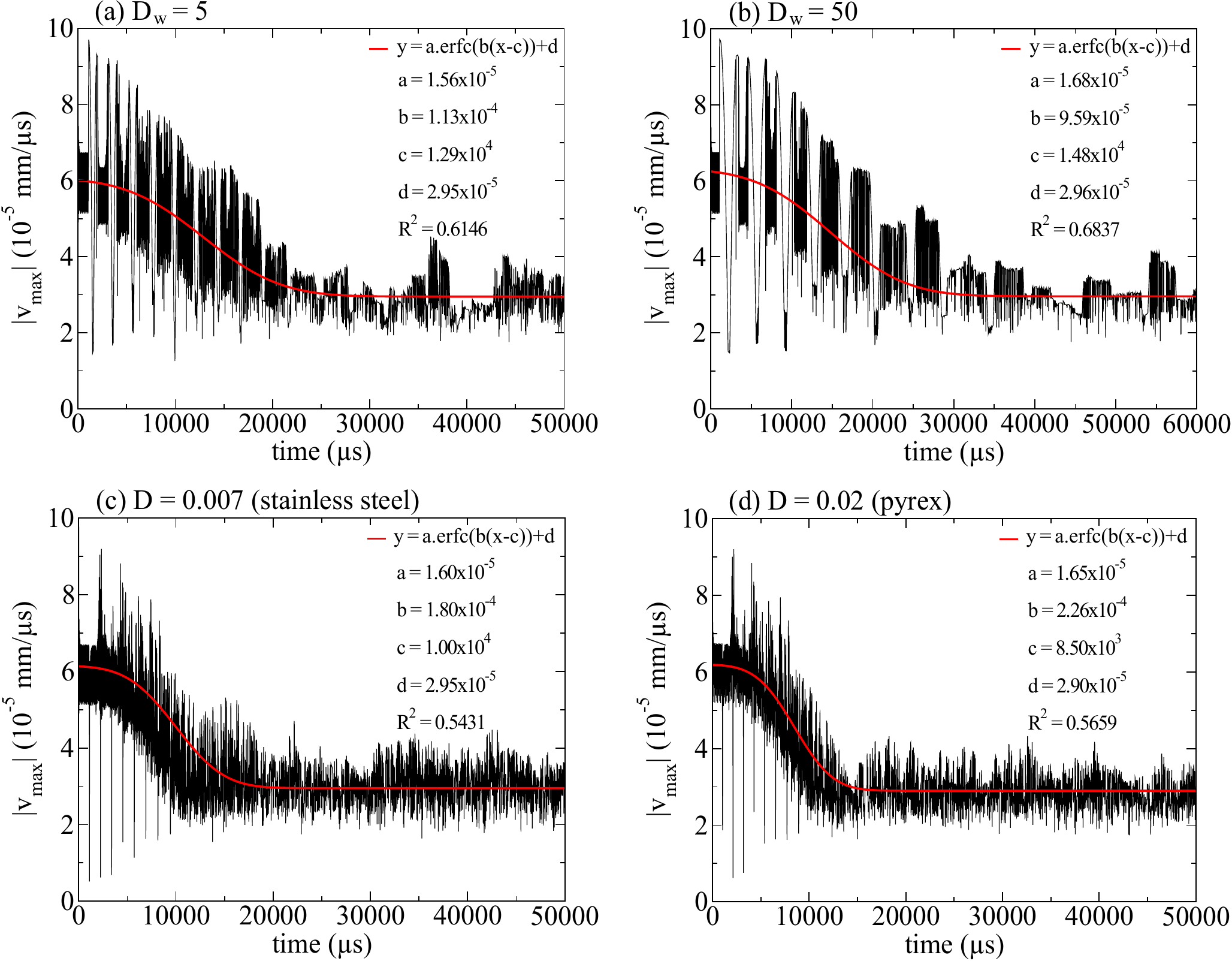}
\caption{(Color online) Speed of the fastest moving grain in the $N=38$ granular chain as a function of time. (a) and (b) correspond to inhomogeneous stainless steel ($D_g=0.007$) granular chains with varying wall softness $D_\textrm{w}$. (c) and (d) correspond to homogeneous ($D_\textrm{w}=D_g=D$) granular chain systems.}
\label{fig:fig3}
\end{figure*}

We present a series of plots in Fig.~\ref{fig:fig3} which depict the absolute value of the velocity of the fastest moving grain in the chain, $|v_{max}|$, as a function of time for several systems. Initially, $|v_{max}|$ oscillates about a maximal value as the ISW travels through the granular chain. Periodically, when the ISW reaches a boundary most of the system's energy converts to stored potential energy in the walls, and $|v_{max}|$ drops to a small value. This happens only briefly in the homogeneous systems since the time it takes for grains to transfer kinetic energy to each other is approximately equal to the duration of rebounding of the ISW off the wall. However, there is an additional time delay in the rebounding of SWs at the soft wall in the inhomogeneous systems, and the interplay between these time scales leads to the emergence of a beating pattern in the soft-wall systems, apparent in Figs.~\ref{fig:fig3}(a) and~\ref{fig:fig3}(b). In this case, the inherent time scale set by the material properties of the walls is not equivalent to the grain-grain energy transfer time, as it is in the homogeneous systems. It is possible that one might be able to exploit this frequency mismatch and beating response by constructing piezoelectric systems with a targeted frequency response.

\begin{figure*}[htp]
\centering
\includegraphics[width=0.9\textwidth]{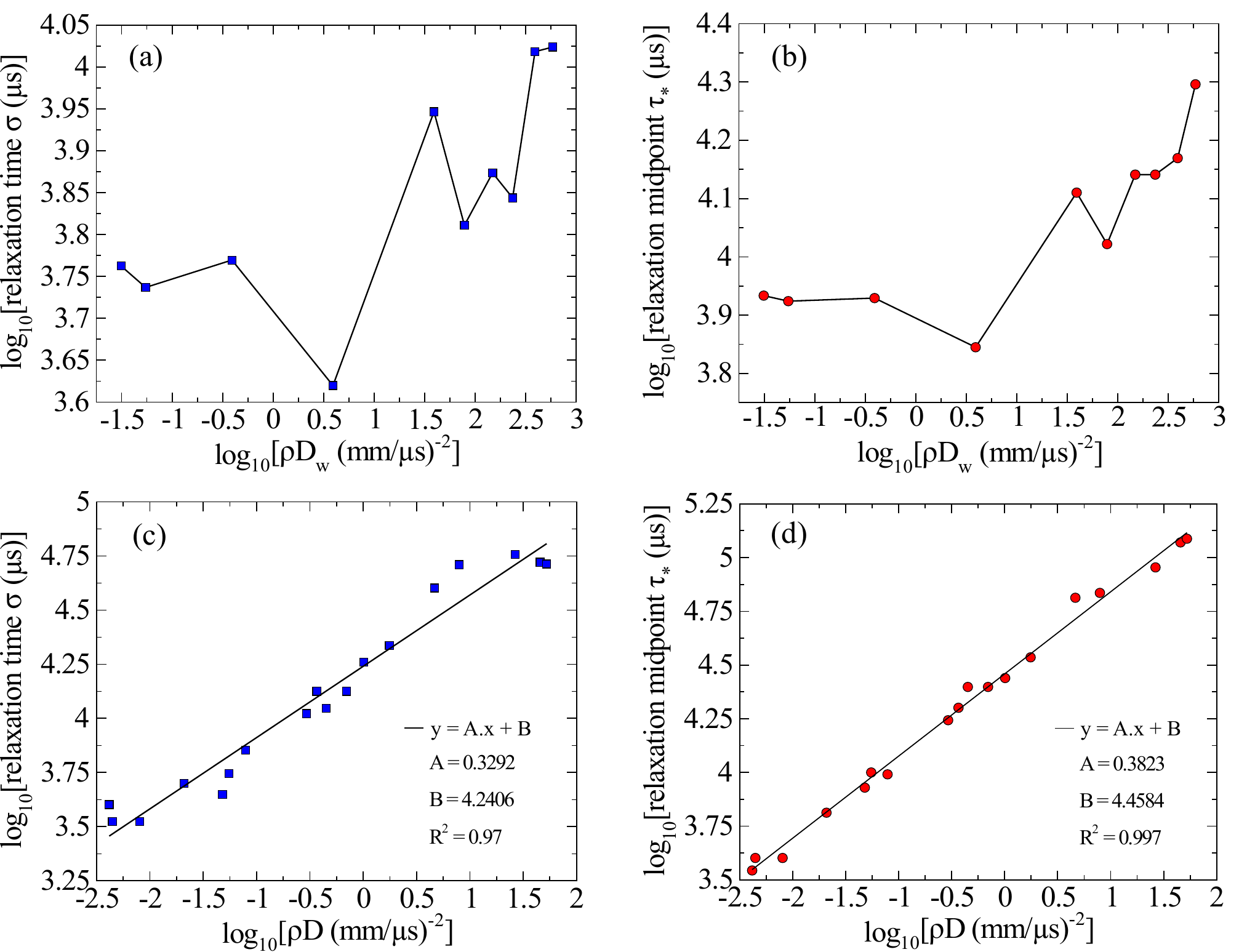}
\caption{(Color online) Results of complementary error function fits to $|v_{max}|(t)$ from Fig.~\ref{fig:fig3}, with logarithmic scales on both axes. Plots (a) and (b) display, respectively, the dependence of the relaxation time $\sigma$, and the midpoint of relaxation to QEQ $\tau_{*}$, on the wall material parameter $\rho D_\textrm{w}$. (c) and (d) illustrate how the relaxation time and midpoint of relaxation, respectively,
depend on the grain material parameter $\rho D$ in homogeneous systems. Solid lines in (c) and (d) illustrate fits to the homogeneous results.}
\label{fig:fig4}
\end{figure*}

Some time after the initial velocity perturbation, $|v_{max}|$ relaxes to noisy oscillations about a much smaller value, denoting the onset of the QEQ phase. One noticeable feature of Figs.~\ref{fig:fig3}(a) and~\ref{fig:fig3}(b) is the delay of the relaxation process introduced by softening the walls. We model this phase transition with a sigmoidal error function of the form:
\begin{equation}
|v_{max}| = v_{\textrm{A}}\left(\mathrm{erfc}\left[\frac{t-\tau_*}{\sigma}\right]\right) + v_{\textrm{B}}.
\label{eq:vmax}
\end{equation}
In this way one can extract the relaxation time $\sigma [\mu\textrm{s}]$, as well as the midpoint of relaxation to QEQ $\tau_* [\mu\textrm{s}]$. A comparison of $\sigma$ and $\tau_*$ are presented in Figs.~\ref{fig:fig4}(a) and~\ref{fig:fig4}(b), respectively, for inhomogeneous systems, and Figs.~\ref{fig:fig4}(c) and~\ref{fig:fig4}(d), respectively, for homogeneous systems. This two-state transition model makes sense for the homogeneous cases, where $|v_{max}|$ exhibits a clear changeover from SW state to the QEQ state, e.g. Figs. 3(c) and 3(d).

However, it is more difficult to implement for the soft-wall systems, where energy is locked-up in the walls as potential energy for extended periods of time, e.g. Figs. 3(a) and 3(b). These gaps in the trend of $|v_{max}|$ prohibit robust fitting of Eq.~\ref{eq:vmax}, and lead to poor goodness of fits ($R^2\sim0.5-0.6$). Unless a better statistical measure of the transition can be found, the exact values of $\sigma$ and $\tau_{*}$ for the inhomogeneous case must be considered imprecise, although the overall trend is qualitatively correct.

Results for the inhomogeneous stainless steel systems, Figs.~\ref{fig:fig4}(a) and~\ref{fig:fig4}(b), illustrate that one can not only slow the rate of relaxation of the system to QEQ beyond its corresponding homogeneous value, but one can also physically delay the onset of the journey from the SW state to the QEQ  state by softening the walls of the system. We reiterate here that ``soft'' is a relative term in these systems, and one can define the walls to be soft only if $D_\textrm{w} > D_g$. The important point is that the walls can be made arbitrarily soft, and thus it should be possible to somewhat tune or delay the transition to QEQ as needed in a real physical system. 

For the homogeneous systems, Figs.~\ref{fig:fig4}(c) and~\ref{fig:fig4}(d), we find a strict power-law dependence of both $\sigma$ and $\tau_*$ upon the parameter $\rho D$, where $\rho$ is the grain/wall density and $D = D_g = D_\textrm{w}$ is the force prefactor, introduced in Eq.~\ref{eq:D}. In particular, one finds that as the grains and walls are simultaneously softened in homogeneous systems, both the relaxation time and onset of relaxation are shifted to higher values. 

To further characterize the granular chains, we considered the fluctuations in kinetic energy from the value predicted by the virial theorem. In particular, the fluctuation in the system's total kinetic energy $\sigma_{K}$ was computed as the relative root mean squared deviation over intervals $m$ which run from time $t_m \to t_m + \tau$:
\begin{equation}
\sigma_{K}(t_m) = \left(\frac{1}{\langle K\rangle_{\textrm{v}}}\right) 
\left(\frac{1}{n_{\tau}}
\sum_{t=t_m}^{t_m+\tau}
\left(K(t) - \langle K\rangle_{\textrm{v}}\right)^2
\right)^{1/2}.
\end{equation}
In the above equation, $\tau$ is the considered time interval (100 $\mu s$ here), $n_{\tau}$ corresponds to the number of data points in this interval, and the mean of the system's kinetic energy 
$\langle K \rangle_{\textrm{v}}$ is a constant equal to the value predicted by the virial theorem, and not the actual average $\langle K \rangle$ of the interval. Since all systems were given the same initial velocity perturbation, the total energy of homogeneous systems depends upon the material properties of the granular chain itself, and $\langle K \rangle_{\textrm{v}}$ is therefore different for each of the homogeneous cases.

\begin{figure*}[htp]
\centering
\includegraphics[scale=0.8]{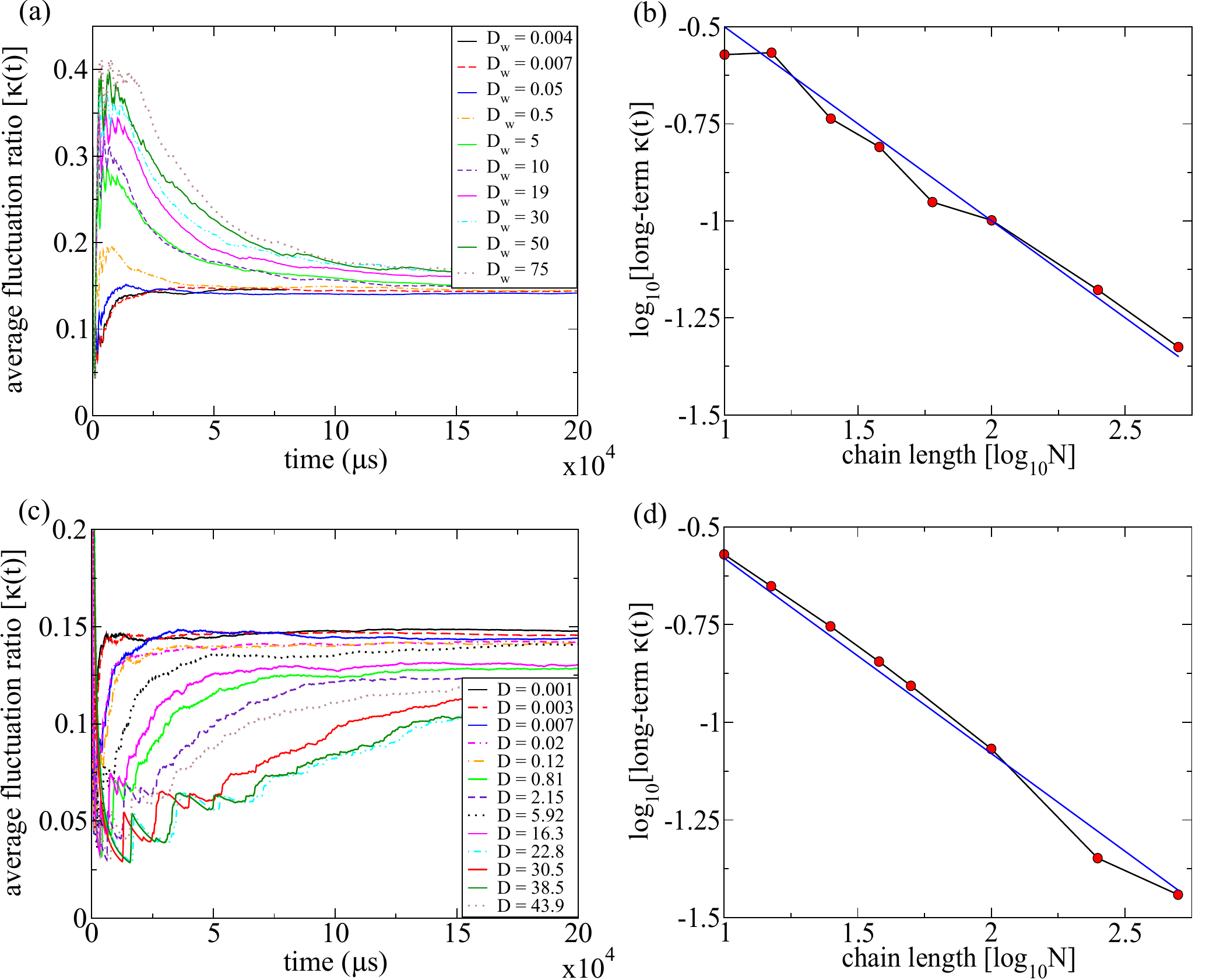}
\caption{(Color online) (a) Average fluctuation in total kinetic energy for $N=38$ stainless steel granular chains with varying wall softness parameter $D_\textrm{w}$. (b) Average long-term fluctuation $\kappa(t \to \infty)$  as a function of chain length $N$ in the stainless steel granular chain with rubber walls ($D_\textrm{w} = 19$). (c) Average long-term kinetic energy fluctuation for the $N=38$ homogeneous chains of varying material parameters. (d) Average long-term fluctuation as a function of system size for homogeneous stainless steel systems. Lines in (b) and (d) have slope of $-1/2$, indicating the approximate dependence $\kappa(t \to \infty) \sim N^{-0.5}$.}
\label{fig:fig5}
\end{figure*}

We compute the average fluctuation in the system's kinetic energy as the average of the intervals $m'$ from the start of the simulation,
\begin{equation}
\kappa(t_M) = \frac{1}{M}\sum_{m'=1}^{M}\sigma_{K} (t_{m'}).
\end{equation}
These results are shown in Fig.~\ref{fig:fig5}(a) for stainless steel chains with varying wall softness, as well as for an array of homogeneous systems in Fig.~\ref{fig:fig5}(c). For the inhomogeneous systems, we see that $\kappa(t)$ exhibits a series of peaks which are modulated by an overall exponential decay at early times, and at much longer times, $\kappa(t)$ relaxes to a long tail with algebraic dependence (we find, e.g., $\kappa(t\to\infty) \sim 1/t$ for the $N=38$ stainless steel chain with $D_\textrm{w}=5$). Deep into QEQ, the average fluctuation in the system's total kinetic energy levels off at a constant value which is insensitive to the wall softness parameter $D_\textrm{w}$, and for chains of length $N=38$ is $\approx 15\%$ of the kinetic energy value predicted by the virial theorem. 

We performed additional simulations of chains of varying length $N$ for homogeneous stainless steel systems, as well as  inhomogeneous chains with stainless steel grains and rubber walls ($D_\textrm{w} = 19$).  The long-term average fluctuation is sensitive to the system size, $\kappa(t\to\infty) \sim N^{-0.5}$ (approximately) for $N \gg 1$ in both cases, as shown in Figs.~\ref{fig:fig5}(b) and 5(d). Thus kinetic energy partitions more evenly for large systems.

\subsection{ \label{sec:longterm} Long-term behaviour}
\begin{figure*}[htp]
\centering
\includegraphics[width=0.9\textwidth]{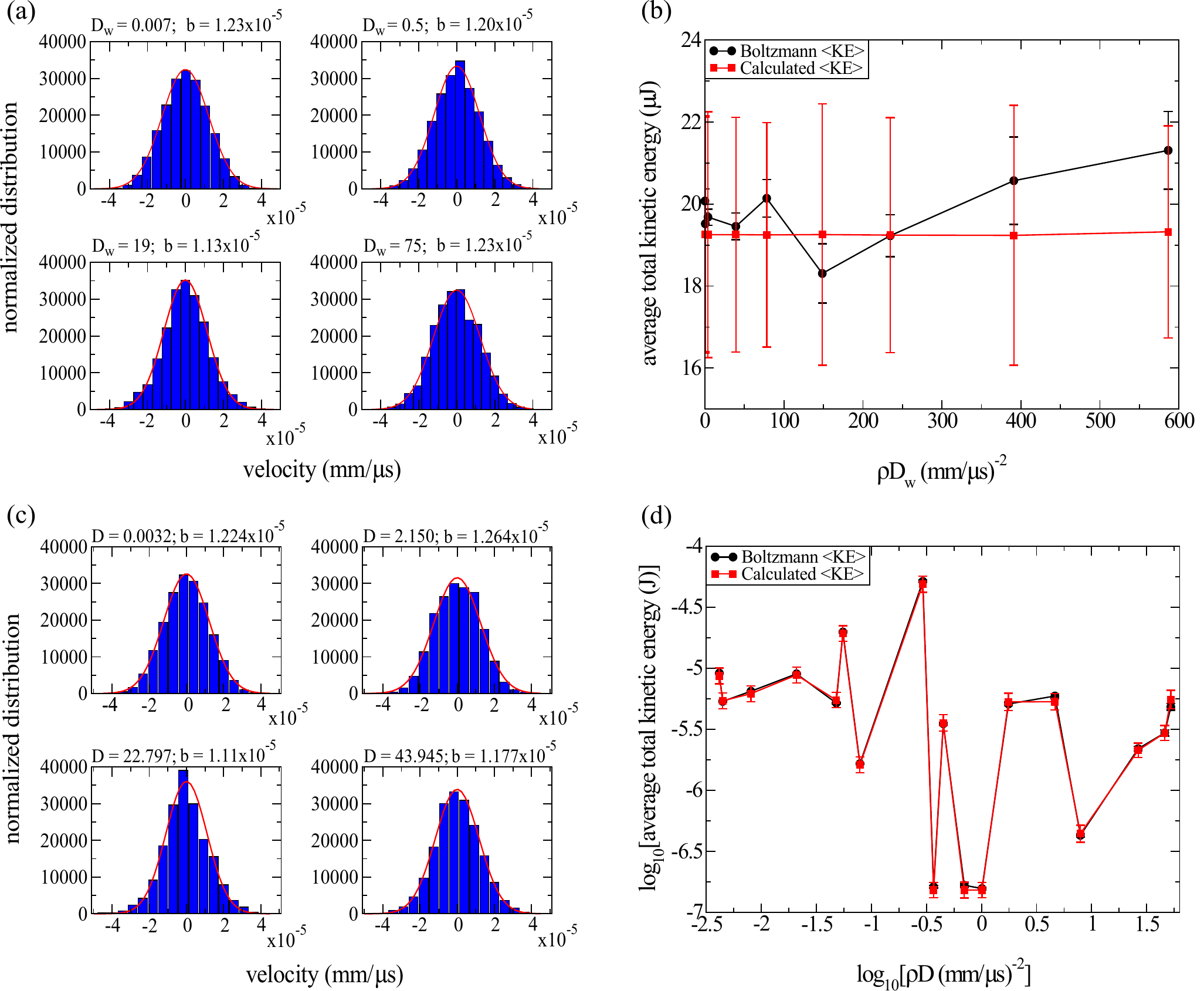}
\caption{(Color online) (a) Maxwell-Boltzman fits to the normalized velocity distributions in the inhomogeneous stainless steel systems with varying wall softness parameter $D_\textrm{w}$. (b) Comparison of average kinetic energy deep into quasi-equilibrium for stainless steel chains of varying wall softness. Circles are the total kinetic energy $K=Nk_B T$ obtained from the Maxwell-Boltzmann temperature, and squares are $K=\langle \sum mv_i^2/2 \rangle$ from the equations of motion. (c) and (d) illustrate the corresponding plots for homogeneous systems. Note that a logarithmic scale has been used in (d). The average kinetic energy is necessarily different for each system since all systems were given the same initial velocity (not energy) perturbation. Thus, we expect to see an agreement between the two curves in (d), which overlap almost perfectly.}
\label{fig:fig6}
\end{figure*}

Deep in the QEQ phase, for this study taken to be the last 10\% of the simulation, an analysis of the grain velocities for both homogeneous and inhomogeneous systems confirms a Maxwell-Boltzmann distribution, as shown in Fig.~\ref{fig:fig6}, substantiating our choice of system size $N$ was not too small. Analysis of the temperature $T$ obtained via Maxwell-Boltzmann fits of the velocity distributions for each system reveals that $T$ is approximately constant with respect to the wall softness parameter $D_\textrm{w}$, as implied by Fig.~\ref{fig:fig6}(b). Moreover, using the temperature from all the stainless steel and soft-wall systems, the total kinetic energy as determined by $N k_B T/ 2$, is in agreement with the calculated average total kinetic energy determined by the average of the grain velocities, e.g. $\langle \sum mv_i^2/2 \rangle$. 

\begin{figure*}[htp]
\centering
\includegraphics[width=0.9\textwidth]{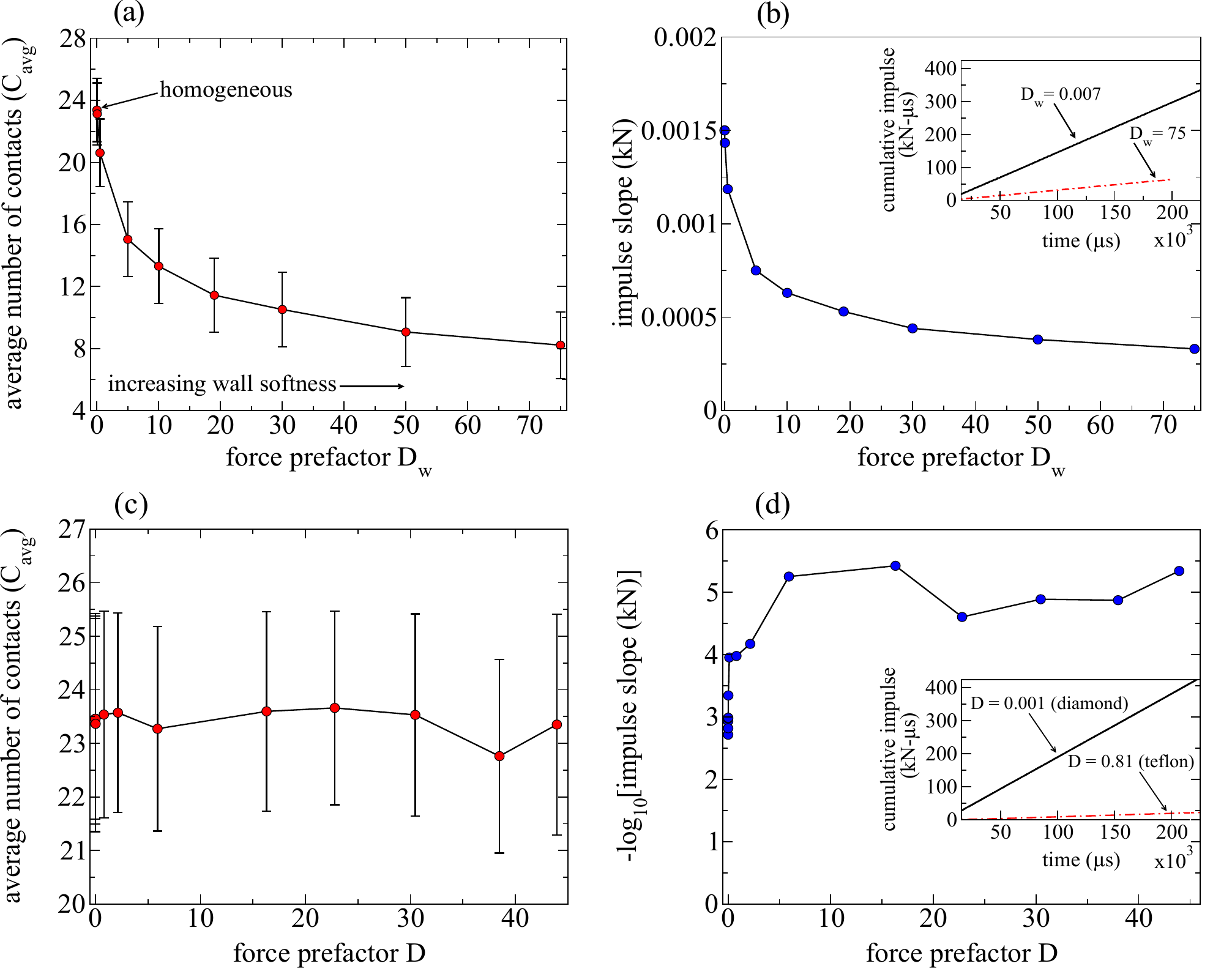}
\caption{(Color online) (a) Average number of grain-grain and grain-wall contacts deep into QEQ for the $N=38$ stainless steel granular chain with varying wall softness, $D_\textrm{w}$. The grains are no longer in contact when $\delta \le 0$. (b) The slope of the cumulative impulse with respect to simulation time, as a function of wall softness. The impulse is perfectly linear in time, shown in the inset for large and small examples of $D_\textrm{w}$. (c) Average number of grain-grain and grain-wall contacts deep into QEQ for the homogeneous $N=38$ granular chains of varying hardness $D$. (d) The slope of the cumulative impulse with respect to $D$. Note the slope is reported as $-\log$ scale, since the very rapid drop to small values as $D>1$. The impulse is again perfectly linear in time, shown in the inset for large and small examples of $D$.}
\label{fig:fig7}
\end{figure*}

Further into the QEQ phase, the kinetic energy is fairly evenly distributed among the grains in our systems, thus any evidence of ``cold'' or ``hot'' spots (such as e.g. in the vicinity of soft walls) have been erased at this point. It turns out that while equipartitioning of energy is expected to be absent in QEQ, it is more difficult to quantify the violation of the equipartition theorem than one might imagine~\cite{Sen2008}. The presence of hot and cold spots from figures similar to Figs.~\ref{fig:fig1} and ~\ref{fig:fig2} have been observed in numerical simulations of granular systems in the QEQ phase~\cite{Avalos2011} and is one of the characteristics which makes this state different from a conventional equilibrium state. We additionally note that the ergodicity of the system deep into QEQ was checked as in Ref.~\cite{Avalos2014}, and both inhomogeneous and homogeneous systems were found to be ergodic. 
Soft walls have the global effect of reducing the average total number of grain-grain contacts in the long-term dynamics of the system. This is illustrated in Fig.~\ref{fig:fig7}(a), which displays the average number of grain-grain and grain-wall contacts deep into QEQ for $N=38$ stainless steel granular chains with varying wall softness parameter $D_\textrm{w}$.  Note that the overall magnitude of the error bars remains relatively constant in Fig.~\ref{fig:fig7}(a), thus the relative fluctuation in the number of contacts increases as a function of $D_\textrm{w}$.

A similar plot is shown in Fig.~\ref{fig:fig7}(c) for homogeneous chains, and it is evident that the average total number of contacts is independent of the material comprising the granular chain. Thus a softening of the boundaries in comparison to grains is required to reduce the average number of contacts. The fact that the end grains spend the whole simulation in close contact with the softer walls opens a lot of space between the walls, which is then distributed between the grains. Paradoxically, the much reduced number of average grain-grain contacts with soft walls does not speed the transition to QEQ, c.f. Figs.~\ref{fig:fig4}(a) and \ref{fig:fig4}(b). This leads one to speculate that while the number of gaps increases, they are more transitory but occur more frequently. 

\begin{figure}[htp]
\centering
\includegraphics[width=0.48\textwidth]{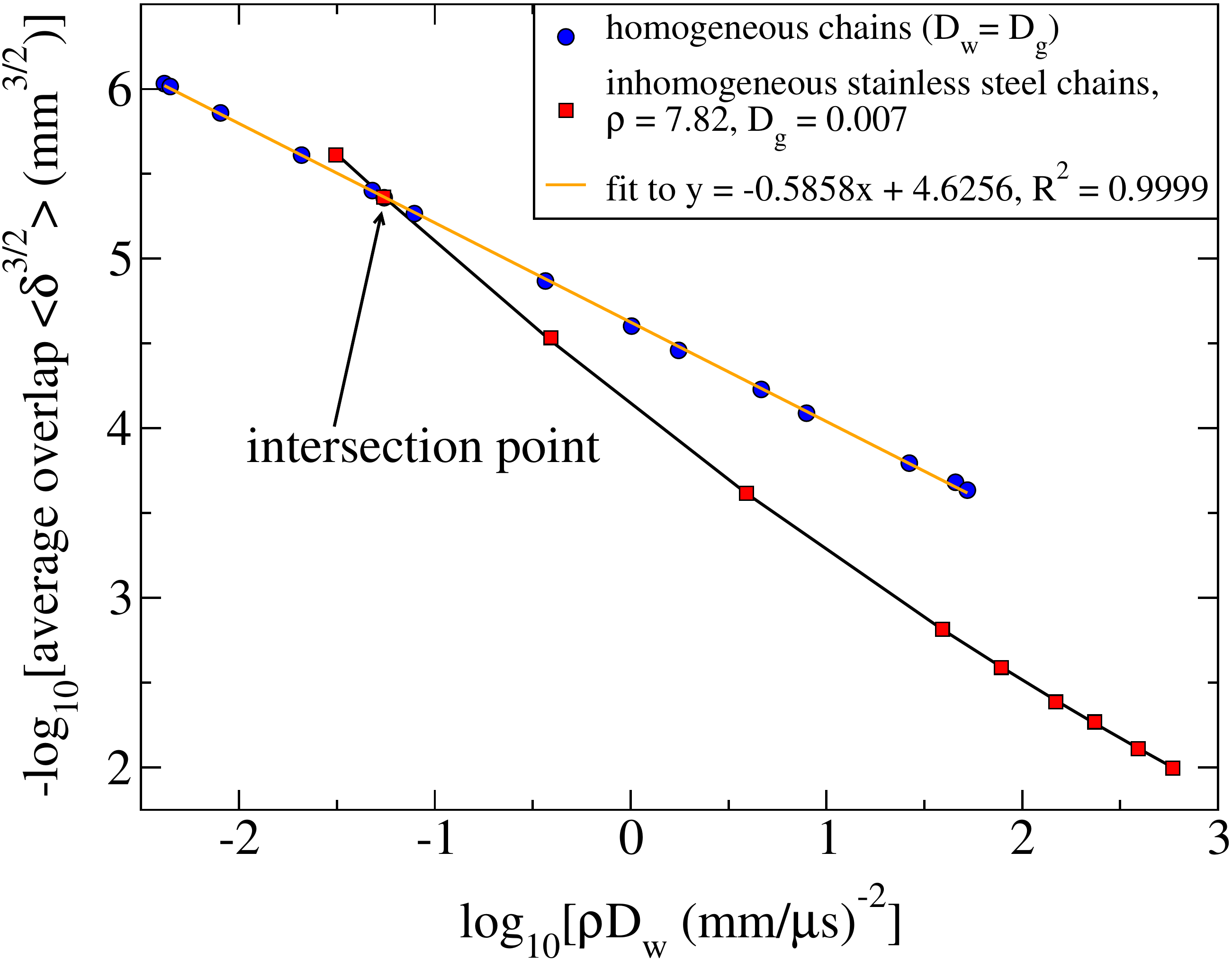}
\caption{(Color online) The average overlap parameter $\langle \delta^{3/2} \rangle$ as a function of material softness $\rho D_\textrm{w}$, where $\rho$ is the grain density and $D_\textrm{w}$ the force prefactor for the end-grain/wall interface. The average end-grain/wall overlap parameter was calculated from linear fits to the cumulative impulse. The intersection point marks the homogeneous stainless steel system. Note that the data is presented on a logarithmic scale for both axes, and the $y$-axis denotes the \textit{negative} logarithm of  
$\langle \delta^{3/2} \rangle$. Thus a smaller value on the $y$-axis corresponds to a \textit{larger} overlap. The single square to the left of the intersection point corresponds to the stainless steel system with harder diamond walls.}
\label{fig:fig8}
\end{figure}

The unbroken contact between the end grains and walls gives rise to a steady force between them,  even as they continue to transfer kinetic energy to neighbouring grains during the reflection of SWs and emission of SSWs at the boundaries.  Despite this sustained force between the end grains and walls, the magnitudes of the forces are increasingly smaller in softer systems. Thus the interplay between increased duration of applied force and decreased magnitude of force leads to an overall smaller time-integrated force (or impulse) in the long term. Over a long time, one sees a linear dependence of the cumulative impulse, $I(t) = \int_0^t F(t')dt'$, between the wall and end grain interfaces with time, as seen in the insets to Figs.~\ref{fig:fig7}(b) and 7(d).

It is evident from Fig.~\ref{fig:fig7}(b) that for a given granular chain, the slope of the impulse curve decreases exponentially as the walls are made softer. Similarly, the slope of the impulse curve is seen to decrease in homogeneous systems as the entire system is softened (i.e. as $D$ increases), see Fig.~\ref{fig:fig7}(d). (Note the $-\log$ scale of Fig.~\ref{fig:fig7}(d).) Inspection of this plot reveals there are anomalies to this rule, suggesting that the density of the grains also plays a role in fully characterizing the ``softness'' of homogeneous systems. Nevertheless, comparisons of Figs.~\ref{fig:fig7}(b) and 7(d) reveal that homogeneous systems ($D_\textrm{w} = D_g = D$) have a much smaller impulse slope at the end grain/wall interface than an inhomogeneous soft-wall system with the same $D_\textrm{w}$ value. 

A linear dependence of the cumulative impulse on time implies that $F(t)$ is a constant of motion in the QEQ phase. Thus the impulse slope is the average force, and the slopes of linear fits to the impulse agree well with the average forces calculated from simulation data deep into QEQ (i.e. from averaging late-time data obtained from directly integrating the equations of motion). Since the force of compression between any two grains is related to the grain-grain overlap via the Hertz potential as $F = (5/2) a \delta^{3/2}$, we can calculate the average overlap between the end grain and adjacent wall through the slope of the impulse curve. Using the expressions from Sec.~\ref{sec:methods}, and taking $R_j \to \infty$ as the radius of the wall, it follows that the average overlap parameter at the end grain/wall interface $\langle \delta^{3/2} \rangle = D_\textrm{w}F_{\textrm{avg}}/\sqrt{6}$ where the grain radius is taken to be $R = 6$~mm. 

We show the average end grain/wall overlap parameter for homogeneous and inhomogeneous systems as a function of $\rho D_\textrm{w}$ in Fig.~\ref{fig:fig8}. Here we include the grain density for all cases to facilitate full characterization of the softness of the homogeneous systems. The data presented in Fig.~\ref{fig:fig8} were calculated from the slopes of linear fits to the impulse curves shown in Figs.~\ref{fig:fig7}(b) and~\ref{fig:fig7}(d). On the log-log plot, see Fig.~\ref{fig:fig8}, the homogeneous systems exhibit a linear relationship with the parameters shown; however, the inhomogeneous systems are well described by a quadratic curve in the figure. We find that deep into the QEQ phase, the system attempts to distribute forces roughly equally among all interfaces in the system. Hence the overlap at the grain/wall interface is larger if one introduces soft walls, which allows this interface to achieve the same force of compression that exists between grain-grain interfaces.

\section{\label{sec:Toy} A toy model}
\begin{figure*}[htp]
\includegraphics[width=0.9\textwidth]{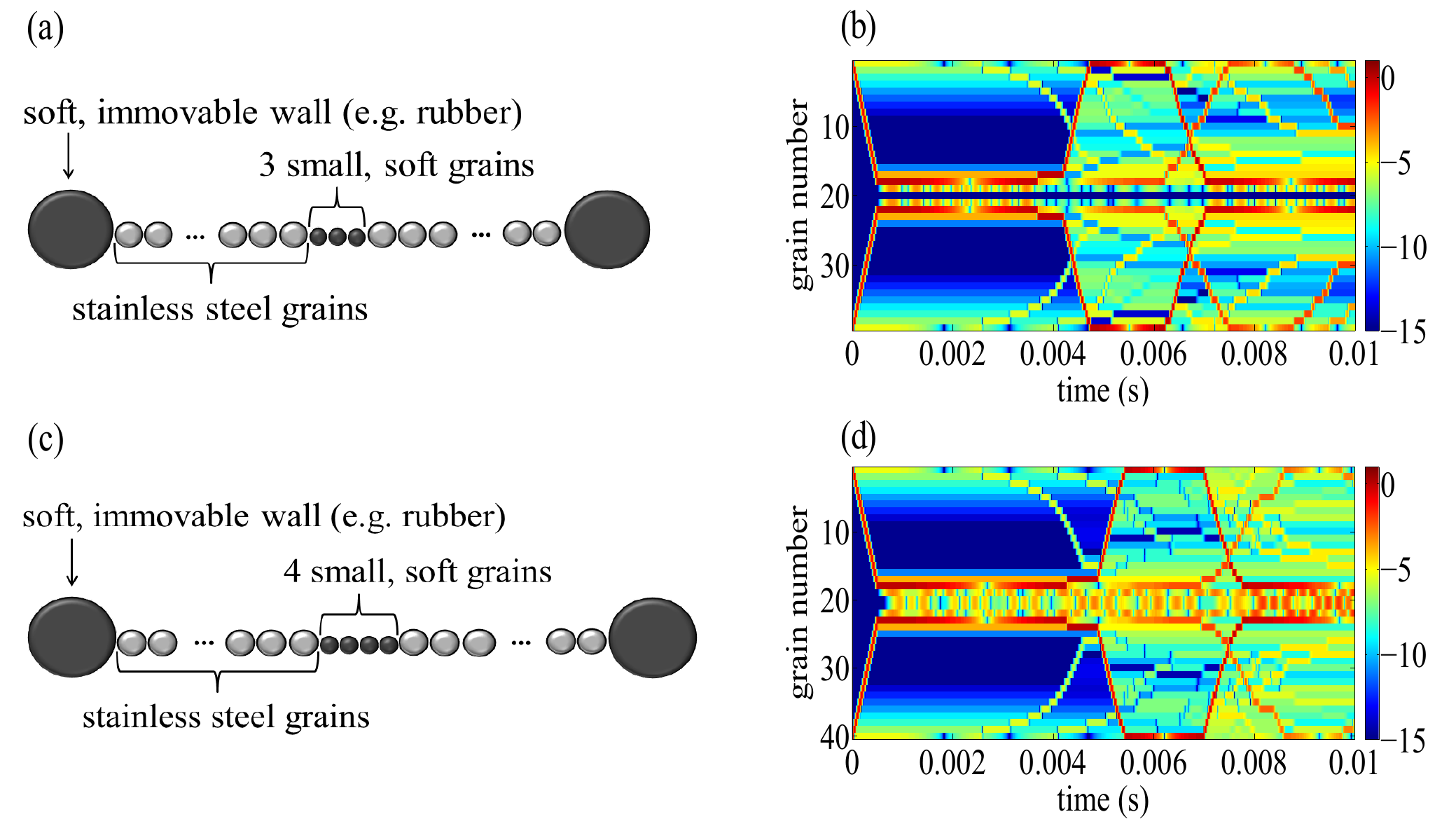}
\caption{(Color online) A toy model for an energy focusing device. Here we have central soft grains composed of a soft material, such as rubber, embedded within a chain of stainless steel grains which are double the radius. To make the system as ``soft'' as possible, we include rubber walls at the edges of the stainless steel chain. (a) Three soft grains in an $N=39$ system, (c) four soft grains in an $N=40$ system. In each system, we initiate two delta-function velocity perturbations at $t=0~\mu$s (one at each edge of the chain), directed toward the center of the chain, and then leave the systems to evolve in time. The corresponding kinetic energy density plots (with energy given on a logarithmic scale and normalized by the energy of the initial perturbation) are displayed, respectively, in (b) and (d).}
\label{fig:fig9}
\end{figure*}

Finally, we discuss a toy model granular system designed for locally trapping energy, and the results of some simulations of this system. In particular, we draw on the results of the previous section, exploiting the ``sticky'' energy propagation associated with the large time scales set by soft materials. We consider in this section a granular chain with soft boundaries and also central soft-grain impurities. The soft boundaries may not be necessary for facilitating the localization of energy at the center of the chain, but are implemented to ensure the system is made as soft as possible. Here we present the results of simulations on soft-wall systems comprised of three or four soft grains located centrally within a chain of hard grains which have double the radius of the impurities, as shown schematically in Figs.~\ref{fig:fig9}(a) and~\ref{fig:fig9}(c). Changing the radius of the impurity does not seem to have a significant effect on the results, though further analysis is required to fully characterize the role of the size of the impurity on the energy localization process. The main result is that there is a remarkable slowing down in the system in the vicinity of the soft impurities. The resulting energy objects are not necessarily so-called discrete breathers, which are generally described as time-periodic, localized, and stable excitations in spatially extended discrete systems. However, the localization of the energy on the impurity behaves effectively like a short-lived breatherbreather, exhibiting an interesting natural frequency response resulting from the mutual repulsion between the grains.

We initiated two delta-function velocity perturbations at $t=0~\mu\mathrm{s}$, one at each edge of the chain, both of magnitude $v_0 = 9.899\times 10^{-5}~\mathrm{mm}/\mu\mathrm{s}$, and directed into the chain. The resulting kinetic energy density plots are displayed in Figs.~\ref{fig:fig9}(b) and \ref{fig:fig9}(d). The initial perturbations develop into SWs which travel toward the center of the chain, meeting precisely at the center due to the symmetry of the system. From Fig.~\ref{fig:fig9}(b), it is evident that the central grain in the odd-numbered $N=39$ chain remains stationary during these collision processes, since the ISWs meet on the central grain itself. Nevertheless, the large time scale associated with the rubber grains causes energy to remain in the vicinity of the soft grains for a noticeably long time. For the even-numbered $N=40$ grain system, the two ISWs meet at the interface between the central grains, thus there is no central cold-spot as there is in the three-soft grain system. This is clear from the kinetic energy density plot for the $N=40$ system (with four soft grains), Fig.~\ref{fig:fig9}(d). This figure also illustrates that energy is localized near the four soft grains in this system for extended periods of time. The temporary energy localization can be attributed to the large time scale associated with the slow energy propagation in the soft impurities, and is also facilitated by the mutual repulsion between grains near the center of the chain.
Also seen in Fig.~\ref{fig:fig9}(b) is the steering of SSWs away from the central soft grains. Around 0.006~s, SSWs approaching the central grains are seen to turn away before settling on grain 15 for a time. Likewise, at 0.009~s, two prominent SSWs stop on adjacent grains. This kinetic energy localization is distinct from potential energy localization on the central grains, since being hard by comparison, these grains will transmit their energy very rapidly if they are in contact with another grain. However, the arrival of a SW and a concomitant opening of a gap in the same direction of the SW prevents the immediate transmission of the energy and allows a temporary localization  of kinetic energy at that grain location.

\begin{figure}[htp]
\includegraphics[width=0.48\textwidth]{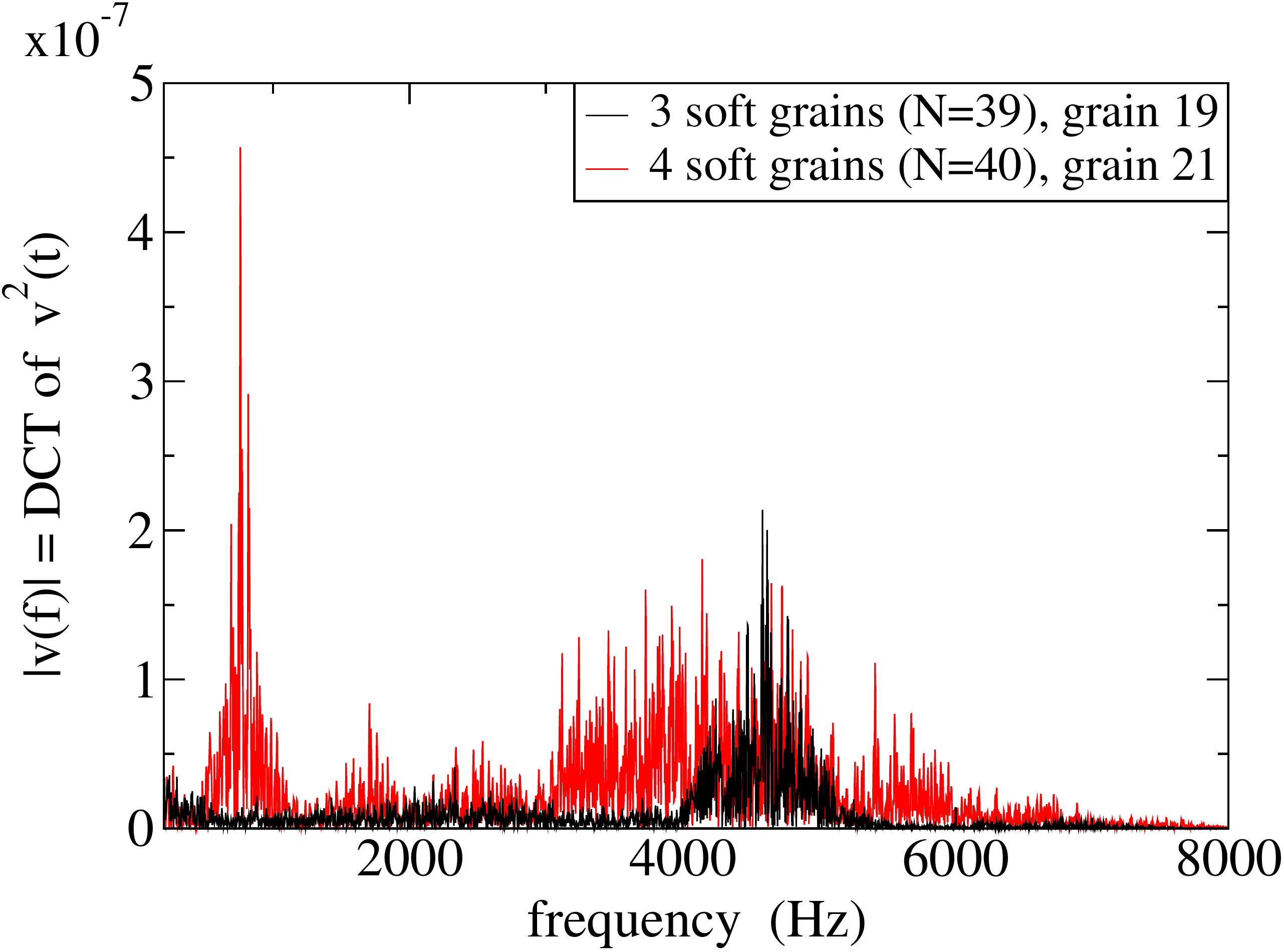}
\caption{(Color online) Discrete cosine transforms of velocities of soft grains in the toy model energy focusing devices. DCTs are performed over the first $100,000~\mu$s, illustrating the presence of a frequency band in the three-soft grain system with $N=39$ grains, and multiple characteristic frequency bands in the four-soft grain system with $N=40$ grains.  Note that the low frequency components ($f<200$~Hz) have been removed from the DCT.} 
\label{fig:fig10}
\end{figure}

To determine how breather-like these objects are, we analyze the frequency profiles via discrete cosine transforms (DCTs) of the square of the grain speed for one of the soft grains. These results are depicted in Fig.~\ref{fig:fig10} for grain $19$ in the three-soft grain system and for grain $21$ in the four-soft grain system. Here we have used the first $100,000~\mu\mathrm{s}$ of the simulation. It is evident that in the three-soft grain system, a single frequency band develops centered between $4000$ and $5000$~Hz, whereas in the four-soft grain system, we see the development of multiple frequency bands: one wide band centered near $4000$~Hz, and a more narrow band around $1000$~Hz. The presence of more than one frequency band is a common characteristic of a breather~\cite{Herbold2009}, suggesting that the four-soft grain system may be more breather-like of the two systems considered here, and may therefore be particularly useful for locally trapping vibrational energy. Thus, our results show that this toy model may be one way to form breathers in granular chains. An extensive discussion on this aspect of the problem is under preparation.    

\section{ \label{sec:con} Conclusions} 
We have performed here a detailed numerical analysis of material properties of granular chains, and the effects of these properties on the dynamics of, and journey to, the QEQ phase. A crucial finding of our study is that soft materials are characterized by an intrinsically large time scale associated with the propagation of energy, which causes a slow down of system dynamics in their vicinity. This slowing down can be exploited for multiple applications. We see in our work the development of ``sticky'' walls in granular chains with soft boundaries, which are particularly useful for energy localization applications. We also presented a toy model that built on the results of the soft-wall systems, and seem to be promising for localizing energy at desired locations in a granular chain, with the resulting energy objects possessing properties similar to those of breathers. Beyond this, there is a natural beating response in the inhomogeneous soft-wall systems due to the interplay between large time scales of the soft boundaries and small time scales of the harder grains. It is possible that this frequency response might be exploited by properly designed piezoelectric systems.  

One major finding of our work is that, since it takes longer for softer materials to rebound the grain-grain compression~\cite{Avalos2014}, incorporating softer materials at the boundaries of the granular chain causes a time delay in the rebounding of SWs, temporarily localizing energy near the wall. This makes the soft walls appear ``sticky'' in dynamical simulations, and one noticeable consequence of this stickiness is the formation of larger SSWs. This was previously experimentally observed by Job et al., who  established that it is possible to increase the energy content of SSWs by softening the walls, at least for $\rho D\approx0.4\to0.008$~\cite{Job2005}. Moreover, since SSWs are produced slowly in the vicinity of the collision region, the soft-wall systems tend to have slightly hotter temperatures near a boundary for measurably long times, at least in dynamical simulations~\cite{Avalos2011}. This was observed recently~\cite{Avalos2014} via the formation of an energy gradient in the early time dynamics of an unloaded granular chain between asymmetric boundaries. 

We conclude  by remarking on the equipartitioning of energy in the granular chain systems. In general, equipartitioning of energy over time is facilitated by the nature of energy exchange in systems, which typically occurs between individual particles. If, on the other hand, interactions in a system do not permit the exchange of energy between individual particles, i.e. if the energy exchange must happen between small groups of particles \cite{Sen2005} then each particle will not end up having more or less the same energy over time. The concept of temperature in such a system is not as well-defined, yet conceivably, the particles may still have a Gaussian distribution of velocities (via Central Limit Theorem) and reach a state that is independent of initial conditions \cite{Sen2005}. While this system would exhibit an equilibrium-like state, the absence of equipartitioning of energy - which manifests itself as large temperature fluctuations \cite{Mohan2005} - makes it distinct from the equilibrium state that is typically encountered in conventional solids, liquids, and gases. This is precisely the nature of the QEQ state in the granular chain when subjected to a single delta function or short time perturbation at initiation. The QEQ state develops a sufficiently long time after an initial perturbation to the system. 

\begin{acknowledgments}
We are grateful to Adam Sokolow, Matthew Westley and Rahul Kashyap for their interest in this work and for their help in various aspects of the calculations. This work was supported by a Vanier Canada Graduate Scholarship from the Natural Sciences and Engineering Research Council. S.S. thanks US Army Research Office for partial support of this research through a Short Term Innovative Research Grant.
\end{acknowledgments}

\appendix*
\section{Material properties}
Table~\ref{table1} lists the material parameters used in simulations, sorted from hard to soft according to $\rho D$ where $\rho$ is the material density. The prefactor between grains $i$ and $i+1$ is $D_{i,i+1}$, which is related to the Young's modulus $Y$ and Poisson's ratio $\sigma$ of the grains by Equation~\ref{eq:D}.

\begin{table*}[hp]
\begin{tabular}{rrrrr}
\hline
$\rho D$ & Young's modulus & Poisson's ratio & Density $\rho$ & Corresponding material \\ 
$(\mathrm{mm}/\mu\mathrm{s})^{-2}$ & ($\mathrm{GPa}$) & & ($\mathrm{mg/mm^3}$) &  \\
\hline
0.00415 & 1220 & 0.2 & 3.52 & Diamond \\
0.00445 & 720 & 0.17 & 2.2 & Quartz \\	                      
0.00805 & 449.65 & 0.207 & 2.52 &  Boron carbide \\	
0.0209 & 249 & 0.18 & 3.58 & Magnesium oxide \\
0.0479 & 67 & 0.2 & 2.23 & Pyrex \\
0.0551 & 193 & 0.305 & 7.82 & Stainless steel \\
0.0786 & 11 & 0.35 & 0.657 & Red oak wood \\
0.293 & 97 & 0.21 & 19.816 & Plutonium \\
0.367 & 0.19 & 0.5 & 0.062 & Polyurethane-II \\
0.450 & 4.1 & 0.39 & 1.45 & PVC \\
0.698 & 0.1 & 0.5 & 0.062 & Polyurethane-III \\
1.01 & 0.069	& 0.5 & 0.062 & Polyurethane-I \\	
1.76 & 1.46 & 0.46 & 2.17 & Teflon \\	
4.644 & 0.55 & 0.46 & 2.16 & Parylene teflon (PTFE) \\
7.91 & 0.032	& 0.25 & 0.18 & Cork \\
26.5 & 0.04 & 0.4334 & 0.87 & Polyolefin elastomer \\
46.2 & 0.03 & 0.48 & 1.2 & Rubber, hard \\
52.4 & 0.05 & 0.49 & 2.3 & Silicone rubber \\	
\hline  
\end{tabular}
\caption{Material properties used in simulations.}\label{table1}
\end{table*}
%

\bibliography{granular}

\end{document}